\documentclass[a4paper,10pt]{article}
\usepackage[affil-it]{authblk}
\usepackage[utf8]{inputenc}
\usepackage{hyperref}  
\usepackage{amsmath,amssymb,lmodern} 
\usepackage{braket}
\usepackage{color}
\usepackage{soul}

\newcommand{\be}{\begin{equation}}
\newcommand{\ee}{\end{equation}}

\begin{document}
\title{On relativistic particle creation in Bose-Einstein condensates}
\author{Carlos Sab{\'\i}n}
\author{Ivette Fuentes}
\affil{School of Mathematical Sciences, University of Nottingham, University Park, Nottingham NG7 2RD, United Kingdom }
\date{}
\maketitle
\begin{abstract}
We show that particle creation of Bogoliubov modes in a Bose-Einstein condensate due to the accelerated motion of the trap is a genuinely relativistic effect. To this end we show that Bogoliubov modes can be described by a time rescaling of the Minkowski metric. A consequence of this is that Rindler transformations are perceived by the phonons as generalised Rindler transformations where the speed of light is replaced by the speed of sound, enhancing particle creation at small velocities. Since the non-relativistic limit of a Rindler transformation is just a Galilean transformation entailing no length contraction or time dilation, we show that the effect vanishes in the non-relativistic limit.
\end{abstract}
\paragraph{Introduction} 

Bose-Einstein condensates (BEC) are natural testbeds for analogue gravity scenarios \cite{analoguereview2011}. By artificially changing the parameters of the condensate, several spacetime metrics such as black holes or expanding universes \cite{megamind} can be simulated while the real spacetime metric remain flat. 

Recently, we and our collaborators have initiated a new avenue of research in which we have shown that changes of the real spacetime metric such as real acceleration or gravitational waves are able to generate observable effects in the BEC \cite{rqm,gravwaves,salelites}, which can be relevant in quantum information space-based experiments and can be used as working principle of ultra sensitive quantum measurement devices. 

Although related, both programmes of research are neatly different. Since a BEC is commonly treated as a non-relativistic system and the relevant scale of speed is provided by the speed of sound -which  is typically as low as a few $mm/s$- it is tempting to think that there could not possibly be relativistic phenomena in these systems. It is the aim of this work however to clearly discriminate between analogue and genuine relativistic effects in BECs and indisputably confirm that the phenomena described in \cite{rqm,gravwaves,salelites} belong to the latter.

In order to build on solid grounds it is obviously necessary to first state what we mean by \textit{relativistic}.  It is a widespread belief that an effect is relativistic only if it occurs at velocities close to the speed of light. However, technological developments have enabled the measurement of a paradigmatic special relativity phenomenon such as time dilation at velocities as low as 10 $m/s$ \cite{wineland}. The accelerated rate at which state-of-the-art quantum metrology devices evolve, promises to achieve unforeseeable levels of accuracy and precision. It thus looks more rigorous to define relativistic effect as something that must be described within a framework consistent with the postulates of Einstein's relativity, regardless the scale of speeds involved in the experiment. That is the case in \cite{wineland}: Lorentz transformations are required to reproduce the experimental results, which cannot be explained by means of Galilean transformations. In this work we will show that this is indeed the case for the generation of particles and mode-mixing described in \cite{rqm,gravwaves,salelites}: they can only be properly described by Rindler transformations- a particular instance of Lorentz transformations- and vanish if Galilean transformations are considered.

\paragraph{The effective metric of Bogoliubov modes}
We start by describing the BEC on a general spacetime metric following references \cite{matt, liberati}.  In the superfluid regime, a BEC  is described by a mean field classical background $\Psi$ plus quantum fluctuations $\hat\Pi$. These fluctuations, for length scales larger than the so-called healing length, behave like a phononic quantum field on a curved metric.  Indeed, 
in a homogenous condensate, the field obeys a massless Klein-Gordon equation 
\begin{equation}\label{eq:dalembertian}
\Box\hat\Pi=0
\end{equation}
 where the d' Alembertian operator 
 \begin{equation}
 \Box=1/\sqrt{-\mathfrak{g}}\,\partial_{a}(\sqrt{-\mathfrak{g}}\mathfrak{g}^{ab}\partial_{b})\label{eq:dalembertian2}
 \end{equation}
depends on an effective spacetime metric $\mathfrak{g}_{ab}$ -with determinant $\mathfrak{g}$- given by  \cite{matt,liberati}
\begin{equation}
\mathfrak{g}_{ab}=\left(\frac{n^2_0\,c_s^{-1}}{\rho_0+p_0}\right)\left[g_{ab}+\left(1-\frac{c_s^2}{c^2}\right)V_aV_b\right].
\end{equation}
The effective metric is a function of the real spacetime metric $g_{ab}$ (that in general may be curved) and   
background mean field properties of the BEC such as the number density $n_0$, the energy density $ \rho_0$, the pressure $p_0$ and the speed of sound 
\begin{equation}\label{eq:speedofsound}
c_s=c\sqrt{\frac{\partial p}{\partial\rho}}.  
\end{equation}
Here $p$ is the total pressure, $\rho$ the total energy density and $V_a$ is the 4-velocity flow on the BEC. This description stems from the theory of fluids in a general relativistic background \cite{matt}, and thus is valid as long as the BEC can be described as a fluid, that is as long as it remains within the quantum hydrodynamic regime \cite{liberati}. 

In the field of analogue gravity, the real spacetime metric is considered to be flat and, therefore, its effects are neglected. Analogue spacetimes are simulated through the artificial manipulation of what we call the analogue gravity metric, 
\begin{equation}
\mathfrak{G}_{ab}=\left(\frac{n^2_0\,c_s^{-1}}{\rho_0+p_0}\right)\left[\left(1-\frac{c_s^2}{c^2}\right)V_aV_b\right].
\end{equation}
Experimentalists change the background parameters of the analogue metric $\mathfrak{G}_{ab}$ to simulate sonic black holes or expanding universes \cite{analoguereview2011,megamind}. Here we are interested solely on the effects of the real spacetime metric. To ensure this, we consider that in the comoving frame 
\begin{equation}\label{eq:velocityflow}
V=(c,0,0,0) 
\end{equation}
and obtain,
\begin{equation}\label{eq:effmetric}
\mathfrak{g}_{ab}= \left(\frac{n^2_0\,c_s^{-1}}{\rho_0+p_0}\right) \left[g_{ab}+ \begin{pmatrix}(c^2-c_s^2)&0&0&0\\0&0&0&0\\0&0&0&0\\0&0&0&0\end{pmatrix}\right]. 
\end{equation}
If the real spacetime metric is flat, then $g_{ab}=\eta_{ab}$, where $\eta_{ab}$ 
\begin{equation}\label{eq:metricplane}
\eta_{ab}= \begin{pmatrix}-c^2&0&0&0\\0&1&0&0\\0&0&1&0\\0&0&0&1\end{pmatrix}.
\end{equation}
$c$ is the speed of light in the vacuum and we consider Minkowski coordinates $(t,x,y,z)$. 
Therefore, the effective metric of the BEC phononic excitations on the flat spacetime metric is given by,
\begin{equation} \label{eq:phonons}
\mathfrak{g}_{ab}= \left(\frac{n^2_0\,c_s^{-1}}{\rho_0+p_0}\right) \begin{pmatrix}-c_s^2&0&0&0\\0&1&0&0\\0&0&1&0\\0&0&0&1\end{pmatrix}.
\end{equation}
Ignoring the conformal factor -which can always be done in 1D or in the case in which is time-independent- we notice that the metric is the flat Minkowski metric with the speed of light being replaced by the speed of sound $c_s$. Notice that considering a rescaled time coordinate:
\begin{equation}\label{eq:changeofcoord}
 t'=\frac{c}{c_s}t\,;\,x'=x \,;\,y'=y \,;\,z'=z
 \end{equation}
 we recover the standard Minkowski metric in Eq. (\ref{eq:metricplane}) from the one in Eq. (\ref{eq:phonons}). This means that the phonons live on a spacetime that is Minkowski however, due to the BEC ground state properties, time flows at a different rate and excitations propagate accordingly. As a result of this, changes in the real spacetime metric are amplified, becoming observable. 

\paragraph{Rindler transformations}

The aim of this section is the analysis of  Rindler transformations of coordinates and, in particular, their effects on the Bogoliubov modes. Rindler coordinates are suitable for uniformly accelerated observers. For the sake of simplicity and without loss of generality, we will consider a 1+1 dimensional spacetime in what follows. Thus, the line element of Minkowski spacetime is:
\begin{equation}\label{eq:linelement}
ds^2=-c^2\,dt^2+dx^2.
\end{equation}
Now, we consider a Rindler transformation:
\begin{equation}\label{eq:rindlertrans}
t=\frac{\chi}{c}\, \operatorname{sinh}\eta;\,\,\, x=\chi\,\operatorname{cosh}\eta.
\end{equation}
An uniformly accelerated observer with proper acceleration $a$ is static in Rindler coordinates, that is she follows a trajectory of constant Rindler position $\chi_0$:
\begin{equation}\label{eq:rindlertray}
\chi_0=\frac{c^2}{a}.
\end{equation}
The proper time $\tau$ of such observer is 
\begin{equation}\label{eq:propertime}
\tau=\frac{\chi_0\,\eta}{c}.
\end{equation}
The line element transforms under the Rindler transformation into:
\begin{equation}\label{eq:rindlerline}
ds^2=-c^2\,d\tau^2+d\chi^2.
\end{equation}
By using Eqs. (\ref{eq:changeofcoord}) and (\ref{eq:rindlertrans}) we can relate the phonon coordinates $(t',x')$ with the Rindler ones $(\eta,\chi)$. It is easy to find that:
\begin{equation}
t'=\frac{\chi}{c_s}\, \operatorname{sinh}\eta;\,\,\,x'=\chi\,\operatorname{cosh}\eta.
\end{equation}
Therefore, Rindler transformations are seen by the phonons as Rindler transformations that depend on $c_s$ instead of $c$. 
The line element of the phonons is conformal to:
\begin{equation}\label{eq:linelementphon}
ds^2=-c_s^2\,dt'^2+dx'^2,
\end{equation}
and it is transformed under a Rindler transformation to:
\begin{equation}\label{eq:rindlerlinephon}
ds^2=-c_s^2\,d\tau'^2+d\chi^2,
\end{equation}
where $\tau'$ is the proper time given by the line element in Eq. ($\ref{eq:linelementphon}$) of an observer following a trajectory $\chi=\chi_0$:
\begin{equation}\label{eq:propertimephon}
\tau'=\int\,d\tau=\int \frac{ds}{c\,\sqrt{-g_{00}}}=\frac{\chi_0\,\eta}{c_s},
\end{equation}
where $g_{00}=\frac{-c_s^2}{c^2}$ is the corresponding element of the metric defined by the line element in Eq. (\ref{eq:linelementphon}):
\begin{equation}
ds^2=g_{\mu\nu}\,dx^{\mu}dx^{\nu}\,,\,x^{\mu}=(c\,t',x').
\end{equation}
The proper acceleration $a$ of an observer with proper time $\tau'$ is:
\begin{equation}\label{eq:properaccphon}
a=|\sqrt{a^{\mu}g_{\mu\nu}a^{\nu}}|=\frac{c_s^2}{\chi_0}\,;\, a^{\mu}=\frac{d^2x^{\mu}}{d^2\tau'}.
\end{equation}
Putting all the above together, we see that the description of Bogoliubov modes, both for inertial and uniformly accelerated observers, is exactly the same as photons but replacing everywhere the speed of light $c$ by the speed of sound $c_s$. In particular, the change in the state of phonons in a cavity due to a sudden change in acceleration is quantified by the Bogoliubov coefficients 
\begin{equation} \label{eq:bogos}
\alpha_{mn}=(\hat{\phi}_{m},\phi_{n})\,,\, \beta_{mn}=-(\hat{\phi}_{n},\phi^{*}_{m})
\end{equation}
which relate the set of inertial modes $\phi$ with the uniformly accelerated ones $\hat{\phi}$ through the Klein-Gordon inner product \cite{birrelldavies}.
The inertial field operators ${a_{k}}$ are transformed into,
\begin{equation}\label{bogotrans}
\hat{a}_{m}=\sum_{n} \bigl(\alpha^{*}_{mn}a_{n}+\beta^{*}_{mn}a^{\dag}_{n}\bigr)\,,
\end{equation}
where $\hat{a}^{\dagger}_k$ and $\hat{a}_k$ are creation and annihilation operators associated to the accelerated mode solutions. Then the coefficients $\alpha_{mn}$ characterise the mode-mixing induced by the transition from inertial to accelerated motion while $\beta_{mn}$ represent particle creation. In the case of photons in a cavity, $\alpha_{mn}$, $\beta_{mn}$ can be computed analytically \cite{alphacentauri} as a series expansion in the parameter $h$, which is:
\begin{equation}\label{eq:h}
h=\frac{a\,L}{c^2},
\end{equation}
where $a$ is the proper acceleration of an observer in the centre of the cavity and $L$ is the constant proper length of the cavity in the Rindler frame,
\begin{equation}\label{eq:length}
L=\chi_R-\chi_L
\end{equation}
 The larger the value of $h$, the larger the mode-mixing and particle creation created by the inertial-to-accelerated transition. Unfortunately, the value of the acceleration required to achieve $h=0.1$ in an optical cavity is $a\simeq 10^{22}\,\operatorname{m/s^2}$.

However, in the case of Bogoilubov modes in a BEC the relevant parameter is:
\begin{equation}\label{eq:hs}
h=\frac{a\,L}{c_s^2}.
\end{equation}
Using typical BEC parameters $L\simeq 10-100 \operatorname{\mu m}$ and $c_s\simeq 1-10 \operatorname{mm/s}$, we find that $L/c_s^2\simeq1-10^4\operatorname{s^2/m}$ and thus sizeable values can be obtained with very small accelerations. This fact can be used to show non-trivial effects of gravity and motion in space-based experiments involving nano satellites \cite{salelites} and to build up an ultrasensitive relativistic quantum accelerometer \cite{rqm}.

\paragraph{Non-relativistic limit: Galilean transformation}

In the last section, we have seen that particle creation in a BEC through acceleration is a consequence of a Rindler transformation, which induces a non-trivial Bogoliubov transformation of the phononic modes. In this section we will consider the non-relativistic limit of the Rindler transformation. 

Using Eqs. (\ref{eq:rindlertray}) and (\ref{eq:propertime}), we can rewrite Eq. (\ref{eq:rindlertrans}) as: 
\begin{equation}\label{eq:rindlertrans2}
t=\frac{\chi}{c}\, \operatorname{sinh}(\frac{a\,\tau}{c});\,\,\, x=\chi\,\operatorname{cosh}(\frac{a\,\tau}{c}).
\end{equation}
We now define:
\begin{equation}\label{eq:velocity}
v=a\,\tau
\end{equation}
and expand Eq.(\ref{eq:rindlertrans2}) as a power series in the small parameter 
\begin{equation}\label{eq:epsilon}
\epsilon= \frac{v}{c}.
\end{equation}
We find:
\begin{equation}\label{eq:rindlernewton}
t=\tau +\mathcal{O}(\epsilon^3)\,,\,x=\chi_0+\frac{a\tau^2}{2}+\mathcal{O}(\epsilon^3),
\end{equation}
that is, a Galilean transformation. Going to the phonon coordinates, we use again Eq. (\ref{eq:changeofcoord}), and we find:
\begin{equation}\label{eq:rindlernewton2}
t'=\tau' +\mathcal{O}(\epsilon^3)\,,\,x'=\chi_0+\frac{a\tau^2}{2}+\mathcal{O}(\epsilon^3),
\end{equation}
which is also a Galilean transformation. These non-relativistic transformations do not generate neither time dilation nor length contraction and thus they induce a trivial Bogoliubov transformation on the phononic modes. In particular, there is no generation of particles.

\paragraph{Conclusions}

We have shown that the generation of particles due to acceleration in a BEC is a genuinely relativistic effect, since it is generated by a Rindler transformation of coordinates. We show that the Bogoliubov modes of the BEC can be described as massless excitations in a spacetime which only differs from the Minkowski spacetime in the fact that $c$ is replaced by the speed of sound $c_s$. This replacement can be understood as a mere rescaling of the time coordinate. As a consequence of this, Rindler transformations act on the Bogoliubov modes as Rindler-like transformations where $c$ is replaced by $c_s$, which amplifies the particle creation at small velocities without altering its relativistic nature. Indeed this is confirmed by noticing that the non-relativistic limit of a Rindler transformation is a Galilean one entailing neither length contraction nor time dilation. Therefore, particle creation vanishes in the non-relativistic limit.

\paragraph{Acknowledgements}
We would like to thank Mehdi Ahmadi, David Edward Bruschi and Nicolai Friis for interesting discussions concerning this work. 

\end{document}